\documentclass[usenatbib]{mn2e}
\usepackage{natbib}
\usepackage{psfig}
\usepackage{graphicx}
\usepackage{amssymb}

\author[McLaughlin et al.]{M. A.~McLaughlin$^{\dagger1,2,3}$, A. G. Lyne$^4$, E. F. Keane$^4$,
 M.~Kramer$^{4,5}$,  
J.~J.~Miller$^1$, \newauthor D. R. Lorimer$^{1,2}$, R.~N.~Manchester$^6$, F.~Camilo$^{7}$ and I.~H.~Stairs$^8$ \\ $^{\dagger}$ Enquiries to:
maura.mclaughlin@mail.wvu.edu \\ $^1$ Department of Physics,
West Virginia University, Morgantown, WV 26506, USA. \\ $^2$ National Radio Astronomy Observatory, Green Bank, WV 24944,USA. \\ $^3$ Alfred P. Sloan Research Fellow \\ $^4$ University of Manchester, Jodrell Bank
Centre for Astrophysics, Alan Turing Building, Oxford Road, Manchester
M13 9PL, UK. \\ $^5$ Max Planck Institut f\"{u}r Radioastronomie, Auf
dem H\"{u}gel 69, 53121 Bonn, Germany.
\\ $^6$ ATNF-CSIRO, P. O. Box 76, Epping  NSW 1710, Australia.\\
$^7$ Columbia Astrophysics Laboratory, Columbia University, 550 W. 120th Street,
New York, NY 10027, USA.\\
$^8$ Department of Physics and Astronomy, University of British Columbia,
6224 Agricultural Road, Vancouver, BC V6T 1Z1, Canada.\\
}  

\date{Accepted by MNRAS on 20 August 2009} 

\title[Timing Observations of RRATs]{Timing Observations of
Rotating Radio Transients}

\begin{document}

\maketitle

\begin{abstract}
We present radio timing measurements of six
rotating radio transient (RRAT) sources discovered in the
Parkes Multibeam Pulsar Survey. These provide four new phase-connected
timing solutions and two updated ones, making a total of seven of the original 11 reported RRATs now
with high-precision rotational and astrometric parameters.
Three of these seven RRATs have magnetic fields greater than $10^{13}$~G, with spin-down
properties similar to those of the magnetars and X-ray detected isolated neutron stars.
Another two of these RRATs have long periods and large characteristic ages, and lie near
the `death-line' for radio pulsar emission. The remaining two RRATs with timing solutions have properties typical of the bulk of the pulsar population. The new solutions offer insights
into what might be responsible for the unusual emission properties. We demonstrate that the
RRATs have significantly longer periods and higher magnetic fields than normal radio pulsars,
and find no correlation with other spin-down  parameters.
 These solutions
also provide precise positions, which will facilitate follow-up studies at high energies, crucial
for relating these sources with other neutron star populations.

\end{abstract}

\begin{keywords}
  stars:neutron -- pulsars: general -- Galaxy: stellar content 
\end{keywords}

\section{Introduction}
\label{intro}

In \citet{mll+06}, we reported the
 discovery of 11  radio-emitting neutron stars 
characterised by repeating dispersed bursts. These ``Rotating Radio
Transients'' (RRATs) have
periods $P$  ranging from 0.7 to 7 seconds,
longer than those of
most normal radio pulsars and similar to those of the populations
of the X-ray detected but seemingly radio-quiet isolated neutron stars \citep[INS; see][]{kaplan08}
 and magnetars \citep[see][]{wt06}.  For
the three RRATs with the highest pulse detection rates, period
derivatives $\dot P$ were measured and reported in the discovery paper. 
Interpreting these $\dot{P}$ values as being due to magnetic dipole braking, they imply
characteristic ages and magnetic field strengths in the general range
of the normal pulsar population. One of these three RRATs, J1819$-$1458, however, has a magnetic field ($5\times10^{13}$~G) in the same range as
the INSs and magnetars.

There have been several suggestions put forward on the nature of this
new class of neutron star. The RRAT emission could be similar to 
 that responsible for the ``giant pulses'' observed from some
pulsars \citep[e.g.][]{kbm+06}. Zhang et al. (2007) \nocite{zgd07} suggest that
the RRATs may be neutron stars near the radio ``death line'' \citep{cr93} or may be
related to ``nulling'' \citep{rr09} radio pulsars.
 Another intriguing possibility is
that the sporadicity of the RRATs is due to the presence of a
circumstellar asteroid belt \citep{li06,cs08} or a
radiation belt as seen in planetary magnetospheres \citep{lm07}.
Alternatively, they may be transient X-ray magnetars, a
 relevant suggestion given the  detections by Camilo
et al. (2006,2007) of transient radio pulsations from two anomalous X-ray
pulsars\nocite{crh+06,crhr07}. A final possibility is that they are similar
objects to PSR~B0656+14, one of three middle-aged pulsars \citep[``The
Three Musketeers'';][]{bt97} from which pulsed
high-energy emission has been detected \citep{dcm+05}.
Weltevrede et al. (2006) \nocite{wsrw06} convincingly show that if PSR~B0656+14 were
more distant, its emission properties would appear similar to those of
the RRATs. Radio polarization measurements of J1819$-$1458 also suggest properties similar to those of normal pulsars \citep{karas}. Determining the reason for the unusual emission of the
RRATs is important since statistical analyses show that they may
 be up to several times more numerous in the Galaxy than the normal radio
pulsars \citep{mll+06,kk08}. Furthermore, \citet{kk08} found  that the Galactic supernova rate may be
insufficient to account for the entire population of neutron stars.
 It is therefore important to know if they evolve to or from other neutron star
populations, or if they are an independent and separate class. Popov et al. (2006) \nocite{ptp06} show that the
inferred birthrate of RRATs is consistent with that of INSs but not
with magnetars.

To answer the above questions, phase-connected timing solutions are imperative.
Periods and period derivatives allow us to compare the spin-down properties of the
RRATs to those of other neutron star populations. Accurate timing positions are also crucial to
facilitate high-energy observations such as the X-ray measurements of J1819$-$1458 which have revealed
a wealth of interesting phenomenology \citep[e.g. absorption features and extended emission cf.][]
{mrg+07,rmg+09}. Due
to their sporadic nature, obtaining phase-connected solutions for the RRATs
is a challenging endeavour and requires significant telescope resources with long observation times, along with densely spaced observing campaigns. Developing efficient methods for timing these
sporadic sources is therefore crucial. This is especially important given the large number
of new RRATs discovered in recent surveys \citep[i.e. five from][one from Hessels et al. 2008, and eight from Keane et al. 2009; note
 we do not include objects in this count which can also be detected through their
time-averaged emission]{dcm+09}.\nocite{hrk+08,klk+09}

With several years of timing data obtained from the Parkes, Lovell and Arecibo telescopes, we have now been able to achieve timing solutions
for four additional RRATs, bringing the total number with solutions up to seven.
 In addition, our continued timing observations of the three original RRATs
have revealed new insights. In another paper \citep{lmk+09}, we report the detection of glitches from J1819$-$1458.
An updated  position for
J1913+1330 is reported in this paper.

In Section~\ref{sec:obs},
 we present the observational data. In Section~\ref{sec:anal}, we describe
our method for timing these sources, and present
 updated solutions for
two RRATs and  new solutions for the other four. In Section~\ref{sec:discuss}, we compare the spin-down properties of the RRATs with
those of other neutron star populations and discuss how these new solutions impact
our understanding of the nature of the RRATs. We offer conclusions and discuss
plans for future work in Section~\ref{sec:conclude}.

\section{Observations}
\label{sec:obs}

 \begin{figure*} \includegraphics[angle=0,width=14cm]{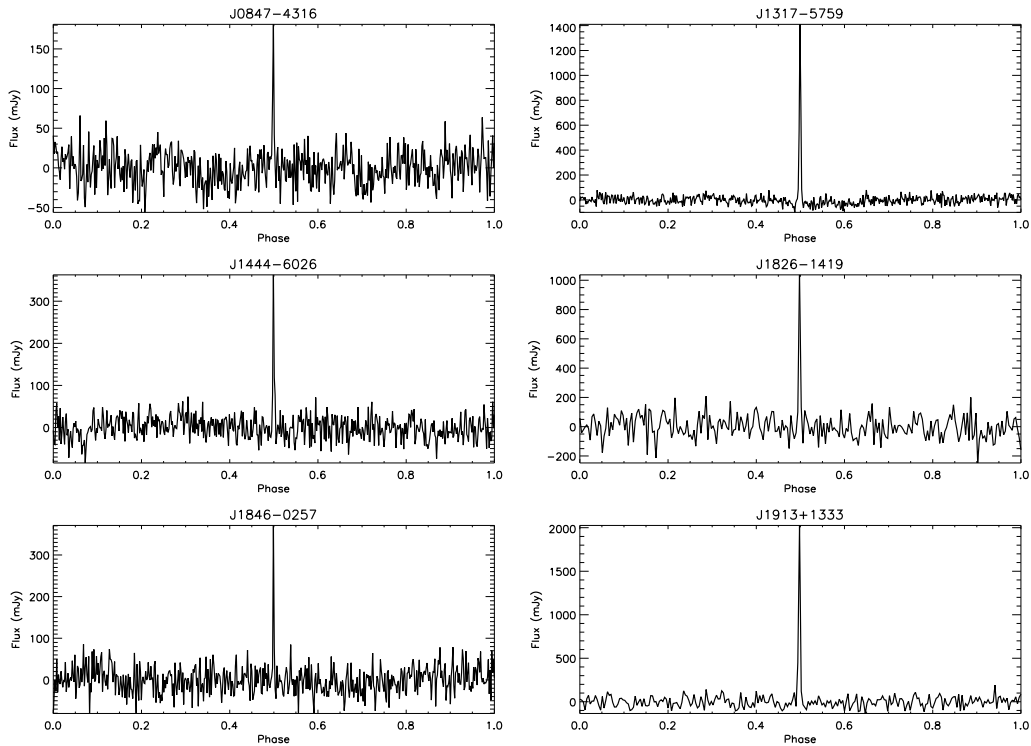}
\caption{ The brightest pulses detected in 1.4-GHz Parkes observations for all six RRATs. There are 512 bins across the pulse profile for each pulse. Flux
densities have been calculated using the radiometer equation \citep[see][]{lk05}, assuming a receiver temperature of 25 K and the appropriate sky temperature at 1400 MHz, scaled from the 408-MHz values  of \citet{hks+81} assuming a spectral index of $-$2.6 \citep{lmop87}. We include a $\sqrt{2/\pi}$ loss in sensitivity for one-bit sampling.
\label{composite}} \end{figure*}

The Parkes Multibeam Pulsar Survey (PMPS) covered the Galactic plane with a 13-beam 1.4-GHz receiver on the 64-m Parkes telescope in NSW, Australia.
 It has now discovered
over 800 new pulsars \citep{mlc+01,kel+09}.
All six of the sources discussed in this paper were discovered by \citet{mll+06} in
a re-analysis of PMPS data taken between MJDs 50842 and 52324 (29 Jan 1998 to 19 Feb 2002). The sources were discovered in 35-min search observations, with between two (for J1444$-$6026) and six (for J1913+1330) pulses detected. Follow-up observations of the new sources with Parkes
began on MJD 52863 (15 Aug 2003) and are on-going, with data up to MJD 54909 (13 Mar 2009) presented here. These observations were spaced at roughly monthly intervals, with some gaps due to telescope scheduling restrictions. All of the
observations were taken using the Parkes analog filterbanks, with 512 0.5-MHz frequency channels sampled every 100~$\mu s$ with 1-bit precision.   Most of the
observations used the central beam of the Multibeam receiver, with a central frequency of 1390~MHz and 256 MHz bandwidth.
However, some observations used the 10$-$50~cm receiver, with a bandwidth of 64 MHz at 685 MHz (50~cm) and a
bandwidth of 768 MHz at 3 GHz (10~cm) and others used the HOH
 receiver, which had a centre frequency of 1.5 GHz
and bandwidth of 576 MHz. In addition, we have regularly observed
 J1913+1330 with the 76-m Lovell telescope at Jodrell Bank,
using a centre frequency of 1402 MHz, and 1-bit sampled with 100-$\mu s$ sampling time and 64 1-MHz channels. We have also observed
this RRAT with the 305-m Arecibo telescope in Arecibo, Puerto Rico using the Wideband Arecibo Pulsar Processor (WAPP) 
 at a centre frequency of 327 MHz with 1024 channels spanning 25 MHz bandwidth and with three-level
 128-$\mu$s sampling.
The six objects for which we present timing solutions are listed in Table~1.
These objects have been observed at between 49 to 71 epochs with observation lengths of 0.5$-$2 hours for each source with Parkes. We have observed J1913+1330 with the Lovell Telescope 
at 62 epochs with observation lengths of 1$-$2 hour. We observed J1913+1330 with the Arecibo telescope
at five epochs with observation lengths of 0.5~hr each.

\begin{table*}
\label{tb:posn}
\caption{
Timing-derived positions, distances of the timing-derived positions
from the center of the discovery-beam positions, longitudes, latitudes, DMs, distances inferred from the \citet{cl02} model, average peak
fluxes and maximum peak fluxes. The numbers in parentheses
 after position and DM are the 1-$\sigma$ errors reported by TEMPO. The distances may be uncertain by roughly 25\%. The peak flux densities were calculated assuming 512 bins across the pulse period. The numbers in parentheses after the average peak flux are the standard deviation.}
\begin{center}\begin{footnotesize} \begin{tabular}{llllrrllll}
\hline Name & R.A. (J2000) & Dec. (J2000) & Offset & {$l$} & {$b$} &  DM & Distance & $S_{1400,{\rm avg}}$ & $S_{1400,{\rm max}}$ \\  & (h~~~m~~~s) & (\degr ~~~\arcmin ~~~\arcsec) & \arcmin & {(\degr)} &{(\degr)} & (pc cm$^{-3}$) & (kpc) & (mJy) & (mJy) \\
\hline
J0847$-$4316 & 08:47:57.33(5)    & $-$43:16:56.8(7) & 4.4 & 263.4   & 0.16 & 292.5(9) & 3.4 & 120(20) & 182 \\
J1317$-$5759 & 13:17:46.29(3) & $-$57:59:30.5(3) & 2.3 & 306.4 & 4.7 & 145.3(3) & 3.0 & 380(200) & 1385 \\
J1444$-$6026 & 14:44:06.02(7)  &  $-$60:26:09.4(4) & 8.9 & 316.4 & $-0.54$ & 367.7(1.4) & 5.5 & 220(70) & 361 \\
J1826$-$1419 & 18:26:42.391(4) & $-$14:19:21.6(3) & 8.0 & 17.4 & $-1.14$ & 160(1)& 3.2 & 520(180) &1048 \\
J1846$-$0257 & 18:46:15.49(4)  & $-$02:57:36.0(1.8) & 2.3  & 29.7  & $-0.20$ & 237(7) & 5.2 & 200(60) & 372 \\
J1913+1330 & 19:13:17.975(8)& +13:30:32.8(1) & 3.4 & 47.4 & 1.38 & 175.64(6) & 5.7 & 460(260) & 2040\\ \hline \end{tabular}
\end{footnotesize}\end{center}\end{table*}

\begin{table*}
\label{tb:prd}
\begin{center}\begin{footnotesize}
\caption{
Periods, period derivatives, MJDs of the epoch used for the period
determination, the average pulse widths at 50\% of the peak, the RMS values of the post-fit timing residual, numbers of pulses included in the timing solution, the
MJD ranges covered, the numbers of epochs and the rates of pulse detection. The numbers in parentheses
 after $P$ and $\dot{P}$ the 1-$\sigma$ errors reported by TEMPO. The numbers in  parentheses
 after $w_{50}$ are the standard deviations. The numbers in parentheses after the total number of observed epochs
are the number of epochs with at least one pulse detected. For J1913+1330, the first/second numbers
are for the Parkes/Lovell telescopes. The rate of pulse detection is much lower for the Lovell telescope due to
	the smaller bandwidth and worse environment for radio frequency interference.} 
\begin{tabular}{llllllllll}
\hline
Name & $P$ & $\dot{P}$ & Epoch & $w_{50}$ & Residual & $N_{\rm p}$ & Data Span  & $N_{\rm e}$  & Rate   \\
 & (s)       & (10$^{-15}$) & (MJD) & (ms) &    (ms)     & & (MJD) & &   (hr$^{-1}$)          \\
\hline
J0847$-$4316 & 5.9774927370(7) & 119.94(2) & 53816 & 27(13)& 11.2 & 138 & 52914$-$54716  & 61(34) & 2.1\\
J1317$-$5759 & 2.64219851320(5)  & 12.560(3) & 53911  & 12(5) & 5.0 & 249 & 53104$-$54717 & 69(60) & 5.2  \\
J1444$-$6026 & 4.7585755679(2) & 18.542(8) & 53893 & 21(7)& 3.2 & 42 & 53104$-$54682 & 71(25)& 0.85\\
J1826$-$1419 & 0.770620171033(7) & 8.7841(2) & 54053 & 2(1.1) & 0.8 & 46 & 53195$-$54909 & 57(18)& 1.0 \\
J1846$-$0257 & 4.4767225398(1) &  160.587(3) & 53039 & 15(6)& 5.4 & 42 & 51298$-$54780 & 49(21)& 1.2\\
J1913+1330 & 0.92339055858(2) & 8.6799(2)  & 53987 & 2(0.7) & 1.1 & 189(159) & 53035$-$54938 & 26(17),62(24) & 13,1.5\\
\hline
\end{tabular}
\end{footnotesize}\end{center}\end{table*}

\begin{figure*} \includegraphics[angle=0,width=12cm]{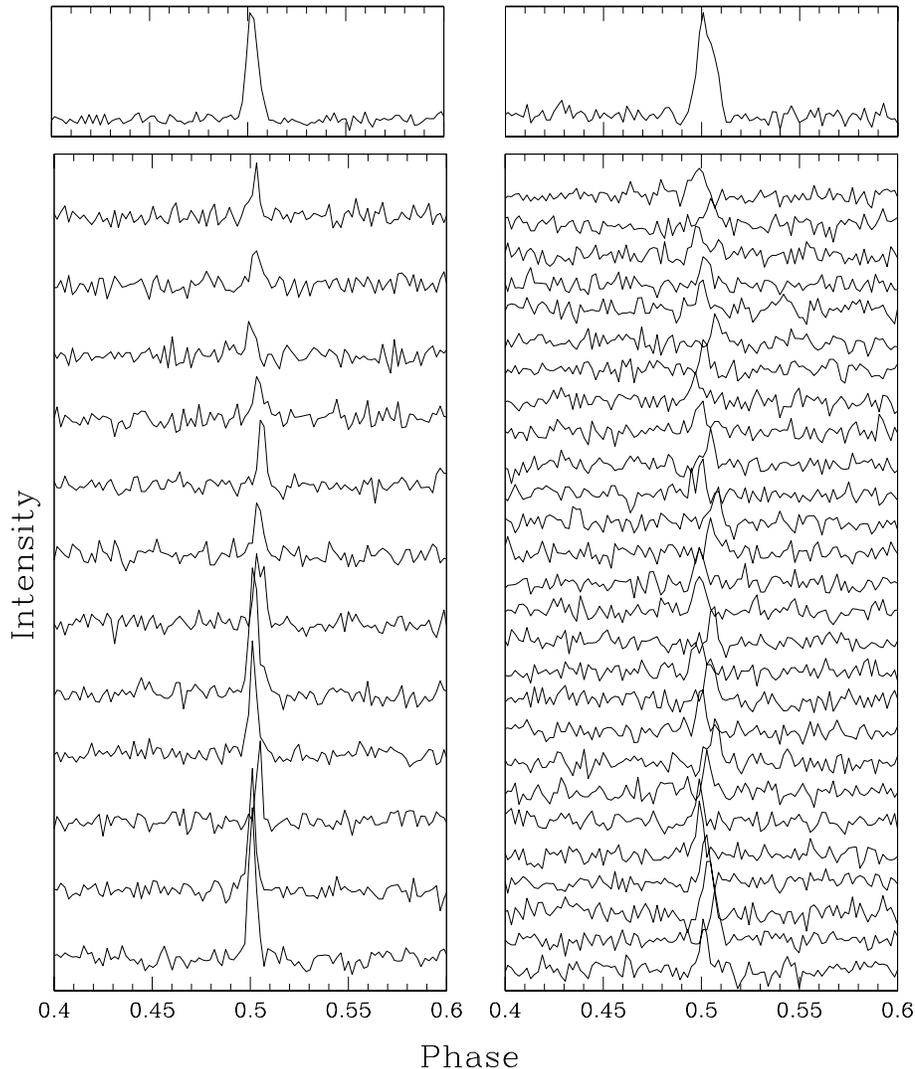}
 \caption{ All of the single pulses detected in a single 1.4-GHz Parkes observation of (left) J1317$-$5759 (MJD 53490) and (right) J1913+1330 (MJD 53156), along with the composite profiles (top) made of all detected single pulses from each observation.
 The observations were one and two hours long, and there are
12 and 27 single pulses for J1317$-$5759 and J1913+1330, respectively. Note that none of these pulses are consecutive,
and the spacing between the pulses varies.
\label{single}}
\end{figure*}
  \begin{figure*} \includegraphics[angle=0,width=14cm]{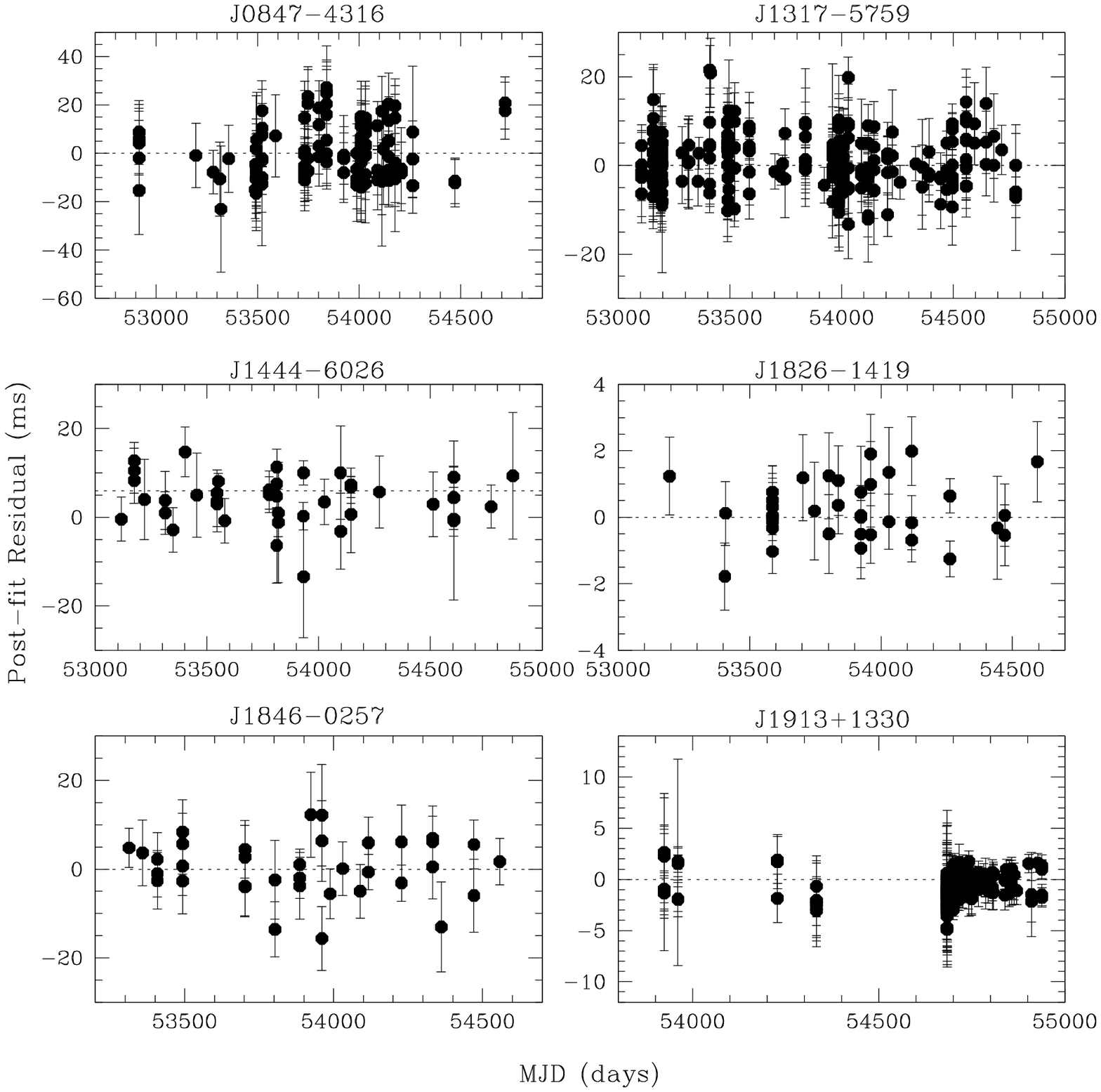} \caption{
Post-fit timing residuals for all six RRATs. The error bars show one-sigma errors on the individual TOAs. All data (i.e. from all frequencies and telescopes) have been included.
\label{resids}} \end{figure*}

 Note that periods for all six objects were published in Mclaughlin et al. (2006).
These periods were measured
 by calculating the differences between the arrival times of all pulses, and finding the greatest
common denominator of these differences. The periods of these RRATs range from 0.77~s (J1826$-$1419) to 5.97~s (J0847$-$4316).
 The number of pulses detected ranges from 42 (for J1444$-$6026 and J1846$-$0257) to 348 (for J1913+1330) for the six objects presented here (Table~1).  Corresponding pulse detection rates range from 0.85~hr$^{-1}$ for J1444$-$6026 to
 13~hr$^{-1}$ for J1913+1330.  The average peak flux densities (assuming 512 bins across the pulse period) range from
120~mJy (for J0847$-$4316) to 520~mJy (for J1826$-$1419).  The maximum peak flux densities detected for each RRAT range from 180~mJy (for J0847$-$4316) to 2040~mJy (for J1913+1330). In Fig.~\ref{composite}, we present the profiles of the brightest pulses detected from each RRAT.
Inspecting the flux densities listed in Table~1 shows that the pulse amplitude distributions of the RRATs vary considerably. A detailed analysis of these distributions will be reported in a companion paper.

\section{Timing Analysis and Results}
\label{sec:anal}

In standard pulsar timing methods,  pulse times-of-arrival (TOAs) are calculated through analysis of integrated profiles formed by summing many ($\sim$ thousands) of
individual pulses modulo the pulse period.
   The sporadic nature  of the RRATs' emission requires us to use single pulses instead of integrated
profiles to calculate TOAs.  
The
first step in  our timing analysis is then pulse detection.  This is done by dedispersing the filterbank data at the dispersion measure (DM) of the RRAT
and at DM of zero, and searching for pulses in both time series above
a 5$\sigma$ threshold in each using the pulsar processing package
SIGPROC\footnote{http://sigproc.sourceforge.net}. Pulses which are brighter at the 
DM of the RRAT are likely to be from the source. We also inspect the pulses visually to be certain of their astrophysical nature. 
For some epochs which have large amounts of RFI, we applied the above procedure but with multiple trial DMs.
 If more than one
pulse is detected within an observation, a second check based on the known period of the source can be made by requiring 
that all pulses have arrival times which differ by integral multiples of the period.
In Table~1, we list the number of epochs for which pulses were detected for all sources. For the sources with more sporadic pulses, like J1444--6026 and J1826--1419, we detect at least one pulse in only $\sim$ 30\% of observations.

To calculate TOAs for normal radio pulsars,
the stable integrated profiles are  cross-correlated with a template profile. 
  The single pulses of all six RRATs discussed in this paper are generally single-peaked. However, the absolute phases of these peaks varies from
pulse to pulse within a window that is typically a few percent of the spin period. In addition, as for normal pulsars, the widths and shapes of the individual pulses of all six vary. In Fig.~\ref{single}, we
show examples of single and integrated profiles for observations of J1317$-$5759 and J1913+1330. Because of this intrinsic
single pulse width variation, common to normal pulsars \citep[e.g.][]{clm+05},
 fitting all single pulses to the same template
 results in large
 systematic errors. We therefore fold the single pulses
with 512 bins across the pulse period to mitigate the effects of noise, and then calculate the TOA as
 the pulse maximum, weighted by the values in the two adjacent bins. The error on the TOA is simply taken as the width of a bin divided by the signal-to-noise of the pulse \citep[see][]{lk05}.

\begin{figure} \includegraphics[angle=270,width=8.5cm]{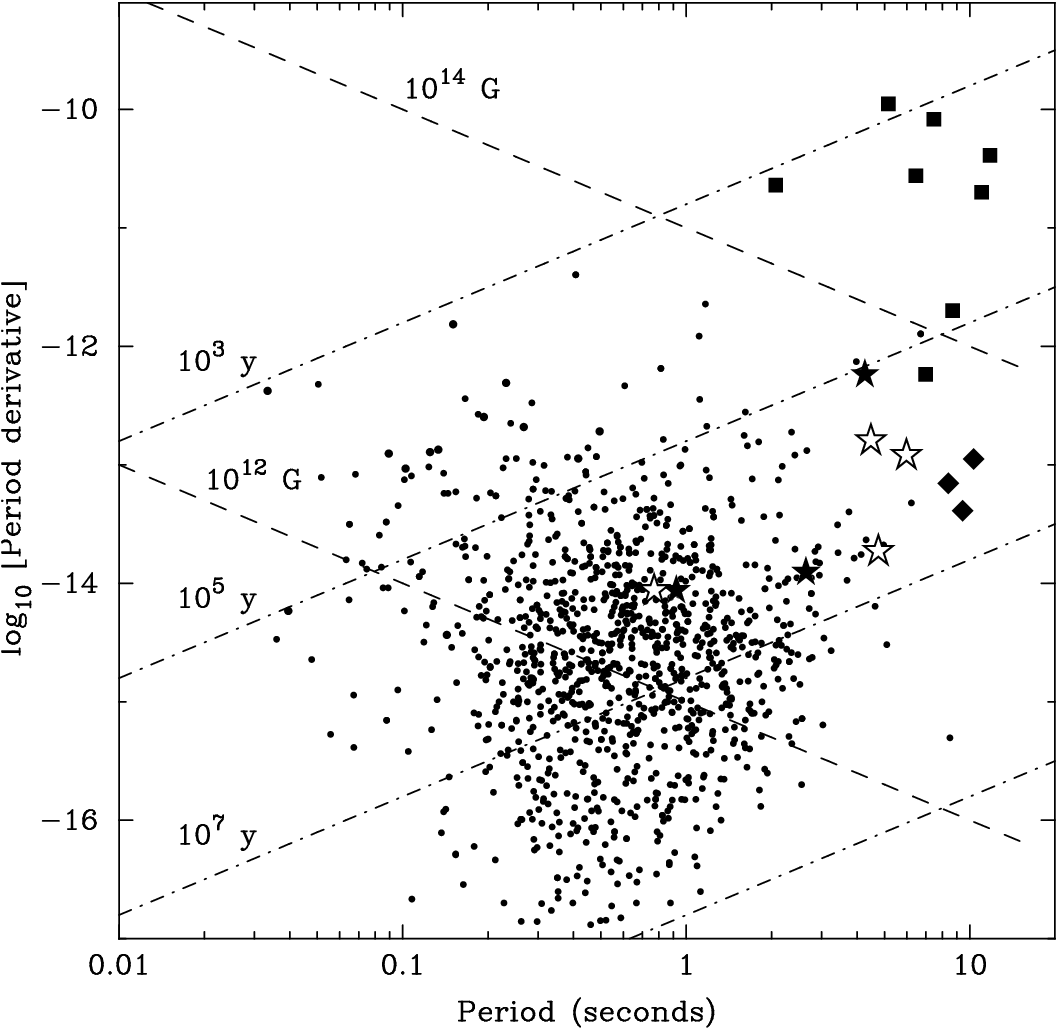} \caption{$P$ vs
$\dot{P}$ for pulsars \citep[dots;][]{mhth05}, magnetars \citep[squares;][]{crhr07},  the original three RRATs with measured periods and period derivatives \citep[red stars;][]{mll+06}, and  the three INS with measured period and period  derivative \citep[diamonds;][]{kk09}. The RRATs with new timing solutions presented in this paper are marked by green stars.   Constant characteristic age and constant inferred surface
 dipole magnetic field strength are indicated by dashed lines. \label{ppdot}}
\end{figure}

The method for fitting a timing model to our TOAs is identical to that for normal pulsars. We use
the TEMPO software package\footnote{http://www.atnf.csiro.au/research/pulsar/tempo} to fit a model incorporating spin period $P$, period derivative $\dot{P}$,
Right Ascension R.A., and Declination Dec. to our data. The results of these fits are shown in Tables~1 and
2. For J0847$-$4316, a frequency second derivative of $4.6(1)\times10^{-25}$~s$^{-3}$ was necessary to fit the data because of a large amount of timing noise.
 Note that our period derivative measurement for J1846$-$0247 is consistent with the upper limit
presented in \citet{aklm08}. 
 There is no evidence for any binary companions to these objects. We see no evidence for glitches,
as seen for J1819--1458 \citep{lmk+09}.
To ensure that the quoted errors in our final model parameters are robust,
we multiply our formal TOA uncertainties by the factor necessary for the reduced chi-squared of our residuals to be equal to one.

In Table~1, we list the offsets of the timing-derived positions from the discovery
positions for the RRATs. For two of the RRATs, the timing-derived position actually lies outside of the 14\arcmin-diameter discovery
beam of the only PMPS detection. Due to the sporadic nature of the emission from these sources, they were not discovered in the PMPS pointings
that were closer to their actual positions. This illustrates the caution with which one must take the discovery positions for these
extreme objects.

For RRATs with detections at multiple centre frequencies, the
DMs in Table~1 were obtained by fitting TOAs at multiple frequencies.
For RRATs with detections only at 1.4~GHz,
DMs were calculated by fitting TOAs in four subbands of the 1.4-GHz bandpass.
The extremely accurate DM for J1913+1330 results from including the sensitive low-frequency observations with the Arecibo telescope in our timing solution. The Arecibo TOAs have formal uncertainties a factor of $\sim$three smaller than the Parkes TOAs. However, despite this increased sensitivity, the source was only detected at one of five epochs with Arecibo due to the short observation time. We therefore
 do not list the overall pulse statistics for the Arecibo observations. 

The post-fit timing residuals for all six RRATs are presented in Fig.~\ref{resids}. The average root-mean-square 
residual ranges from 0.8~ms (for J1826$-$1419) to 11.2~ms (for J0847$-$4316), or roughly 1$-$2\% of the pulse period. 
These are roughly equal to the widths of the composite
pulses for these objects, listed in Table~2. For the six RRATs discussed in this paper, we see no evidence for the 
`banding' structure seen in the J1819$-$1358 residuals
\citep{ezy+08,lmk+09}, and the composite profiles of these RRATs appear to be single.
 The pronounced non-uniformity in the residual spacing for J1913+1330 in Fig.~\ref{resids} 
is due primarily to gaps in observing coverage. For the other RRATs,
any non-uniformity is likely to arise from intrinsic variability
in the rate of pulse emission.
This time variability of burst rates will be discussed in more detail in a follow-up paper.

\section{Discussion}
\label{sec:discuss}

\citet{kk08} found that the Galactic supernova rate was insufficient to account for the entire population of
neutron stars, including radio pulsars, INS, magnetars and the central compact objects. Evolution among the
various types of neutron stars could go a long way towards solving this problem. It is therefore important to
compare the spin-down properties of the RRATs with those of these other objects.
In Table~3, for all of the six objects, we list the inferred surface
dipole magnetic field, the characteristic age, 
the spin-down energy-loss rate, and the magnetic field strength at the light cylinder.
In Fig.~\ref{ppdot}, we place the RRATs on a $P-\dot{P}$ diagram with other populations of neutron stars.
The spin-down properties of J1826$-$1419 and J1913+1330 are remarkably similar to each other 
and place them solidly within the normal
radio pulsar population. This would support models which suggest that the RRATs are simply normal pulsars at the tail end of the (largely unexplored) intermittency distribution \citep{wsrw06}, which attribute the RRATs' unusual emission
to external factors \citep[e.g.][]{cs08,li06} or which theorise that the emission is similar to that of normal
nulling pulsars \citep[e.g.][]{zgd07}.
Two other RRATs, J1317$-$5759 and J1444$-$6026, have long spin periods and large characteristic
ages and are nearing the `death line' for radio pulsars \citep{cr93}.
 These properties suggest that these
RRATs are older pulsars whose radio emission is slowly turning off \citep[e.g.][]{zgd07}.
Two other RRATs for which we have recently obtained solutions,
J0847$-$4316 and J1846$-$0257, have high magnetic fields and are in a region of $P-\dot{P}$ space devoid of radio pulsars.
The proximity of these objects to the X-ray detected but radio-quiet INS suggests that
 these sources may be transition objects between normal radio pulsars and INS,
an idea supported by the population analysis of \citet{kk08}.
It should  be noted that INS may not be an intrinsically radio-quiet population.
Kondratiev et al. (2009)\nocite{kml+09} show that the radio non-detections of INS can easily be
attributed to beaming effects.
 The remaining RRAT, J1819$-$1458, for which a detailed analysis is reported in \citet{lmk+09}, has a high magnetic field and spin-down parameters similar to both some high magnetic field radio pulsars and magnetars.

\begin{table}
\label{tb:deriv}
\begin{center}\begin{footnotesize}
\caption{
Base-10 logarithms of the derived parameters
characteristic age, surface dipole magnetic
field strength,  rotational energy loss rate and
magnetic field strength at the light cylinder. See \citet{lk05} for definitions of these parameters.}
\begin{tabular}{llllr}
\hline
{Name} &
{$\log[\tau_c]$} &
{$\log[B]$} &
{$\log[\dot{E}]$} &
{$\log[B_{\rm LC}]$}\\
 & (yr) & (G) & (erg~s$^{-1}$) & (G) \\
\hline
J0847$-$4316 & 5.9 & 13.4 & 31.3 & 0.1 \\
J1317$-$5759 & 6.5 & 12.8 & 31.4 & 0.5  \\
J1444$-$6026 & 6.6 & 13.0 & 30.8 & $-$0.1 \\
J1826$-$1419 & 6.1 & 12.4 & 32.9 & 1.7 \\
J1846$-$0257 & 5.6 & 13.4 & 31.8 & 0.4 \\
J1913+1330 & 6.2 & 12.4 & 32.6 & 1.5 \\
\hline
\end{tabular}
\end{footnotesize}\end{center}\end{table}

\begin{figure*}
\includegraphics[angle=0,width=14cm]{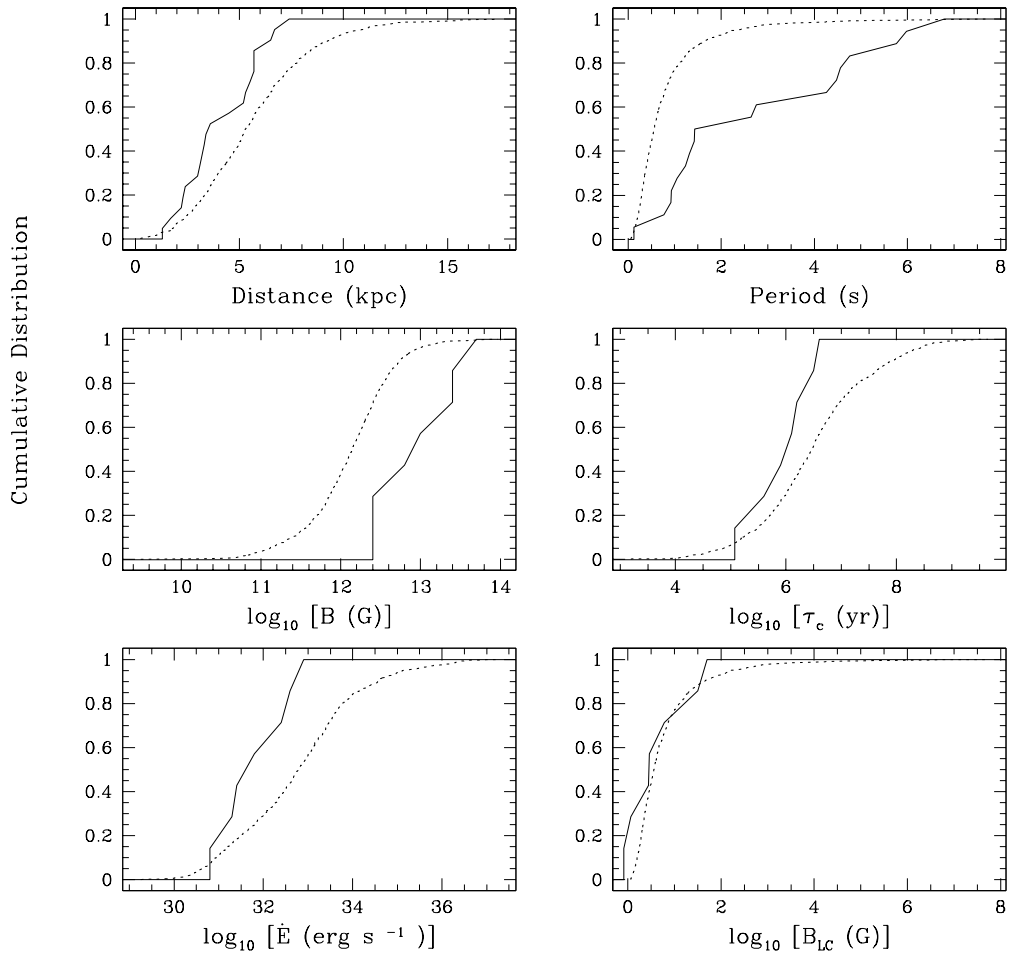}
\caption{Cumulative probability distributions for distance, period, magnetic field, characteristic age, spin-down energy loss rate and magnetic field at the light cylinder for RRATs (solid lines) and all PMPS-detected non-recycled pulsars (dashed lines). The RRAT sample includes all PMPS-detected RRATs (10 from \citet{mll+06} and eight (all but J1841$-$14, which can be detected through its time-averaged emission) from \citet{klk+09} for $P$, 11 from \citet{mll+06} and ten (again excluding J1841$-$14) from \citet{klk+09} for  $d$, and six from this paper and one from \citet{lmk+09} for the other parameters). The PMPS sample includes  1004 pulsars (for $P$ and $d$) and 984 pulsars (for other parameters). Note that while the difference between the two CDFs seems smaller for $d$ than for some other parameters, there are more measured distances and therefore this difference is more significant.
\label{hists}}
\end{figure*}

To test whether the distributions of the derived parameters of the RRATs are consistent with those of the normal pulsar population, we performed a Kolmogorov-Smirnoff \citep[KS, see, e.g.][]{pftv86} test on the  distributions of RRATs and all of the pulsars
detected in the PMPS (see Fig.~\ref{hists}). We use the more uniform PMPS sample, as opposed to all radio pulsars, to mitigate selection effects. For the RRATs distribution, we include all PMPS-detected RRATs 
\citep{mll+06,klk+09}. In Table~4, we list $\cal{P}_{\rm KS}$, or the probability that the two distributions are
drawn from the same parent distribution. We also list $\cal{P}_{\rm ran}$, or the probability of getting a $\cal{P}_{\rm KS}$ as low or lower,
as determined by simulations of 100,000 trials where we randomly select values from the PMPS sample.

For the period distribution, we include all PMPS-detected RRATs with
measured periods. This includes 10 RRATs from \citet{mll+06} and eight (all but J1841$-$14) from \citet{klk+09}.
while for the derived parameters we include the six RRATs discussed in this paper and J1819$-$1458.
 The probabilities that the
  periods and magnetic fields of the RRATs and normal pulsars are drawn
from the same parent distributions are small ($<10^{-3}$),
with the RRATs having longer periods and higher magnetic fields. This suggests that
long period neutron stars with higher magnetic fields are more likely to manifest themselves as RRATs than are
shorter period or lower magnetic field neutron stars.

We believe that the effect is real and not due to a bias against short period objects. Following \citet{mc03}, we
can show that a
single-pulse search will be superior to a periodicity search for periods between $T_{\rm obs} g^2/4 < P < T_{\rm obs} g$,
where $T_{\rm obs}$ is the observation length and
$g$ is the fraction of neutron star rotations where a pulse is emitted.
For $g = 0.001$ and $T_{\rm obs}$ = 35 min (i.e. the length of the PMPS survey observations), this
 period range is
0.5~ms to 126 seconds. Therefore, for a pulse emission rate as high as one per 1000 pulses (i.e. similar to J1913+1330),
the single-pulse
search will result in higher signal-to-noise ratios than a periodicity search for any reasonable periods. For lower
pulse emission rates of $g = 10^{-4}$ (i.e. similar to J1826$-$1419), the single-pulse search will result in
higher signal-to-noise than periodicity search for an even broader range of periods. It is only when we approach much higher pulse emission rates
(i.e. $g$ close to 0.01) that selection effects against short-period objects become important.
We therefore believe that shorter period objects with similar rates of pulse emission to these six sources would
also have appeared as RRATs in the PMPS. As there is no bias against period derivative in this survey, we
infer that the relationship between RRAT behaviour and both period and magnetic field is robust.
The characteristic age, energy loss rate, and magnetic field at the light cylinder distributions of normal pulsars and RRATs are consistent.
This suggests that the RRATs emission behaviour is not determined solely by any of these
quantities.
In Table~1, we list distances inferred from the best-fit DMs using the \citet{cl02} model
for Galactic electron density.
We found that the probability that the PMPS-detected  RRATs distribution (including 11 RRATs from \citet{mll+06} and ten RRATs from \citet{klk+09}) and the PMPS distribution are drawn from
the same parent distribution is $4\times10^{-2}$, with the RRAT population being slightly more nearby.
However, due to the significant selection effects involved in distance distributions it is difficult to draw
conclusions from this weak correlation.

\begin{table}
\label{tb:ks}
\begin{center}\begin{footnotesize}
\caption{
Measured or derived parameter, the probability that the RRATs and all PMPS-detected pulsars
are drawn from the same parent distribution, and the probability of measuring a  $\cal{P}_{\rm KS}$
as low or lower in our simulations.}
\begin{tabular}{lll}
\hline
Parameter & $\cal{P}_{\rm KS}$ & $\cal{P}_{\rm ran}$\\
\hline
$d$ & $4\times10^{-2}$  & $4\times10^{-2}$  \\
$P$ & $7\times10^{-7}$  & $< 1\times10^{-7}$\\
$B$ & $8\times10^{-4}$ & $4\times10^{-4}$\\
$\tau_{c}$ & 0.3 & 0.3 \\
$\dot{E}$ & 0.1 & 0.1 \\
B$_{\rm LC}$ & 0.5 & 0.5 \\
\hline
\end{tabular}
\end{footnotesize}\end{center}\end{table}

Pulsars which emit giant pulses have been shown \citep[e.g.][]{kbmo05} to have values of
magnetic field at the light cylinder $B_{\rm LC}$ greater than 10$^{5}$~G
and spin-down rotational energy loss $\dot{E}$ values
greater than 10$^{35}$~erg~s$^{-1}$.
The values of $B_{\rm LC}$ and $\dot{E}$ for all of the RRATs are many
orders of magnitude below these values, suggesting that the RRAT's emission mechanism
is very different.
The pulses of the RRATs are also wider than the widths of pulsar giant pulses,
which have widths of much less than 1~ms \citep[e.g.][]{btk08}.
They range from 2--30 ms, with duty cycles ranging from 0.21\% (for J1913+1333) to 0.45\% (for J0847$-$4316 and J1317$-$5759).

\section{Conclusions and Future Work}
\label{sec:conclude}
 
We have presented a method for timing pulsars through their individual pulses, and have applied this method to RRAT sources originally discovered in the Parkes Multibeam Pulsar Survey. Our work results in four new timing solutions, resulting in seven of 11 of these RRATs having phase-connected solutions. 
 We find that the period and magnetic field distributions of RRATs and normal pulsars are different, and that this difference cannot be explained by selection effects. This suggests that the unusual emission behaviour of the RRATs may be related to their long periods and high magnetic fields. More phase-connected solutions are necessary to
determine any evolutionary relationships between these objects and the long-period, high-magnetic-field populations of magnetars and X-ray-detected isolated neutron stars. The high-energy observations enabled by the new timing positions presented in this paper will be crucial for testing these relationships. 

There remain four RRATs from the original Parkes Multibeam Pulsar Survey analysis for which we do not yet have phase-connected solutions. One of these, J1839$-$01,
has only been detected once despite multiple attempts. The remaining three RRATs have pulses that are too sporadic and
 weak to be timed with the Parkes telescope and are the subjects of intense low-frequency Green Bank Telescope and Arecibo observing campaigns. These campaigns, and the properties of the RRATs
discussed in this paper at low frequencies, will be the subject of a follow-up paper.

\section*{Acknowledgments}

We thank all on the Parkes Multibeam Survey team for assistance with the Parkes radio observations.
We are grateful to Aris Karastergiou for useful comments on the manuscript.
MAM, JJM and DRL are supported by a WV EPSCoR grant.
EK acknowledges the support of a Marie-Curie EST Fellowship with the FP6 Network ``ESTRELA'' under contract number MEST-CT-2005-19669.

\bibliography{journals,psrrefs,modrefs}
\bibliographystyle{mn2e}

\end{document}